\documentclass[fleqn,10pt]{wlscirep}
\usepackage{graphicx}
\usepackage{amssymb}
\usepackage{epstopdf}
\usepackage{url}
\usepackage{subfigure}
\usepackage{lineno}

\title{Predicting Urban Innovation from the  Workforce Mobility Network in US}

\author[1]{Moreno Bonaventura}
\author[2]{Luca Maria Aiello}
\author[2*,3]{Daniele Quercia}
\author[1,4,5]{Vito Latora}
\affil[1]{School of Mathematical Sciences, Queen Mary University of London, London, E14NS, United Kingdom}
\affil[2]{Nokia Bell Labs, Cambridge, CB30FA, United Kingdom}
\affil[3]{CUSP, King's College London, WC2R2LS, United Kingdom}
\affil[4]{Dipartimento di Fisica e Astronomia, Universit\`a di Catania and INFN, 95123 Catania (Italy)}
\affil[5]{The Alan Turing Institute, The British Library NW12DB, London  (UK)}

\affil[*]{quercia@cantab.net}


\begin{abstract} %max 200 words
While great emphasis has been placed on the role of social
interactions as driver of innovation growth, very few empirical
studies have explicitly investigated the impact of social
network structures on the innovation performance of cities. Past research has mostly explored scaling laws of
socio-economic outputs of cities as determined by, for example,  the  single
predictor of population. Here, by drawing on a publicly available dataset of the startup ecosystem, we build the first Workforce Mobility Network among US
metropolitan areas. We found that node centrality computed on this network accounts
for most of the variability observed in cities' innovation performance
and significantly outperforms other predictors such as population size or density,  suggesting that policies and initiatives aiming at sustaining innovation
processes might benefit from fostering
professional networks alongside other economic or systemic incentives.  As opposed to previous approaches powered by census data, our
model can be updated in real-time upon open
databases, opening up new opportunities  both for researchers in a variety of disciplines to study urban economies in new ways, and for practitioners to design tools for monitoring such economies in real-time. 

\end{abstract}
\begin{document}

\flushbottom
\maketitle
\thispagestyle{empty}

\section*{Introduction}

%SITUATION
Over the last two decades, developed and developing countries alike
have witnessed a radical transformation in the nature and dynamics of
their innovation processes. A major factor that has triggered this change
is the emergence of new entrepreneurial ecosystems centered on
high-growth startups. In the United States, startups account for the
majority of new job creations~\cite{decker2014role} and have rapidly expanded 
not only in size but also geographically by creating distributed
innovation centers~\cite{acs2008employment}. Abundant empirical
evidence supports the idea that young and innovative firms guarantee
the long-term growth of cities and sustain the economic life by
creating wealth and new jobs also in related
industries~\cite{mumford1961city,hall1998cities,glaeser2010urban,bos2013gazelles,haltiwanger2013creates,weins2014importance}. 

\begin{figure}
   \centering
   \includegraphics[width=4in]{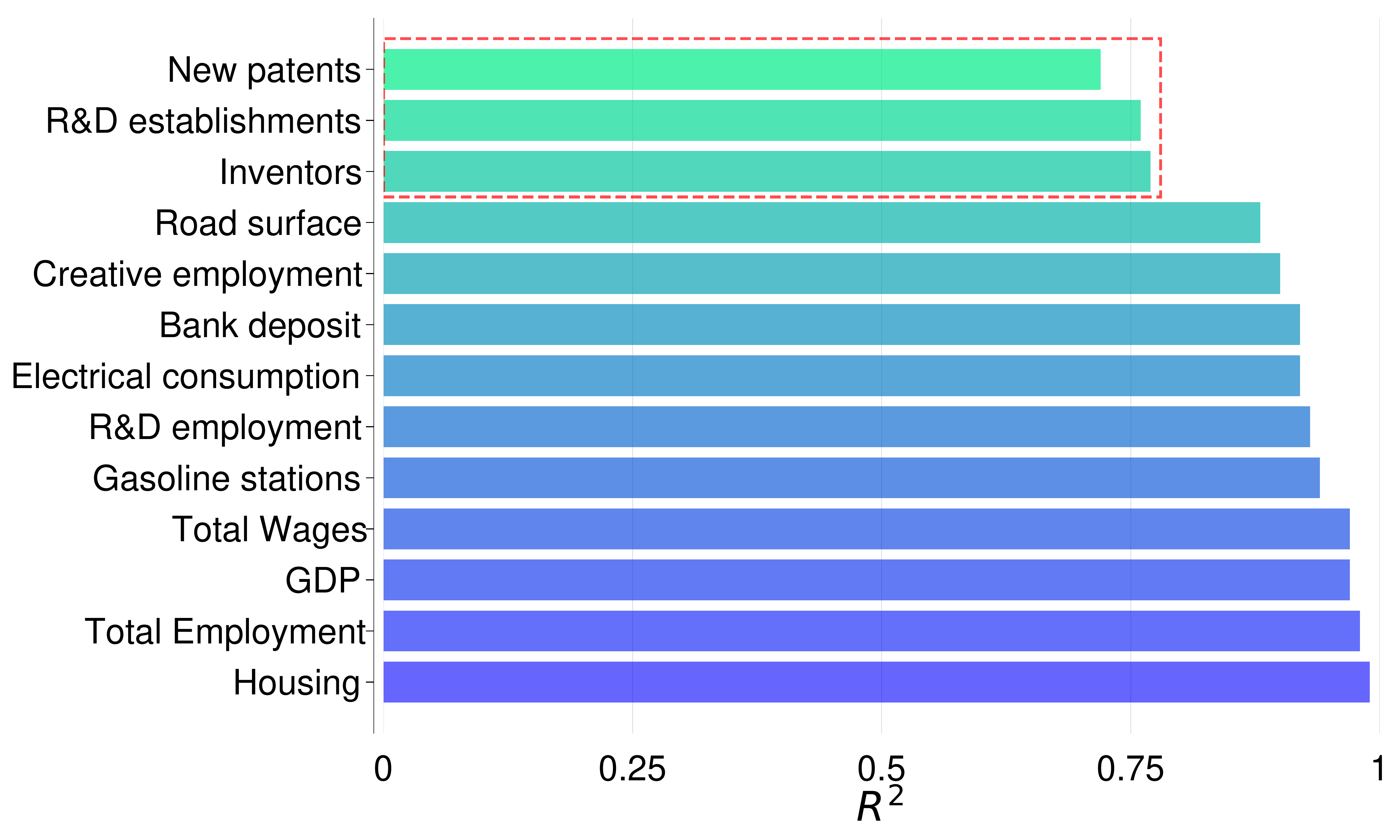} 
   \caption{Correlation coefficients between population size and
     various socio-economic outputs, adapted from L. Bettencourt
     et. al.~\cite{bettencourt2007growth}. Data suggest that
     innovation-related indicators (marked in red) are less correlated with population
     than other socio-economic outputs. These results are computed with data covering the [1997,2003] period.}
   \label{fig:bettencourt_bars}
\end{figure}

%COMPLICATION
Researchers have tried to shed light on early indicators of success in
modern innovation environments. In the attempt of building baseline
models to predict innovation in cities, past efforts have mainly
focused on  predicting  a wide range of
socio-economic indicators of wealth (e.g., GDP, employment, housing and 
infrastructures) and a range of \textit{innovation} indicators
(e.g., abundance of young firms, number of patents
granted)~\cite{bettencourt2007invention,bettencourt2007growth,arcaute2015constructing} solely based on population size or density. These studies have shown that population size alone is able to reliably predict -- with a coefficient of determination $R^2$ for linear
regression in the $[0.88,0.99]$ range  -- several socio-economic
outputs of cities including income, electrical consumption, total
wages, and employment. Yet, the correlations  between population characteristics and  outputs associated with \emph{innovation} processes
such as number of granted patents $(R^2=0.72)$, number of inventors
$(R^2=0.76)$, and R\&D establishments $(R^2=0.77)$ are not equally strong. In fact,
innovation-related indicators report the smallest correlation
coefficients among all the other
variables~\cite{bettencourt2007growth}
(Figure~\ref{fig:bettencourt_bars}).

This discrepancy points to three main limitations of prediction models
solely based on demographic variables. First, by treating geographical
areas as isolated entities, such models overlook the role of social interactions, yet  well-established urban theories\cite{jacobs1970economy} and qualitative \cite{saxenian1996regional} and quantitative findings in economics \cite{glaeser2011triumph} have repeatedly shown that  a dense and dynamic web of interactions among
specialized workers, entrepreneurs, and investors -- also referred to
as the ``thickness of the market'' -- plays a pivotal role  in driving idea recombination,
innovation generation, and ultimately economic
growth~\cite{jacobs1961death,glaeser2001measuring,moretti2012new}. Second, these past models do not account for the fact that cities grow through the attraction of highly talented individuals
(also called `the creative class'~\cite{florida2005cities}), and  the creative outputs from such individuals have been recently found to  explain
super-linear urban scaling~\cite{Keuschniggeaav0042}.  Finally, the
life cycle of a modern innovative startup -- its birth, growth,
acquisition, and extinction -- is much faster than the time frames within which past models' inputs (e.g., demographic data) and outputs (e.g., patenting rates) are typically defined.

Previous research has provided evidence that simple scaling 
laws of population miss evolutionary dynamics that are key to 
explain many city-level processes~\cite{depersin2018global}, and that the application of tools from statistical physics to a variety of 
spatial networks allows for a more accurate description of such complex 
dynamics~\cite{lammer2006scaling,barthelemy2016structure,kirkley2018betweenness,barbosa2018human,barthelemy19statistical}.
However, constrained by limited data availability, only a few empirical
studies have attempted to investigate the impact of different types of social network
structures on economic growth and innovation performance of
cities~\cite{bettencourt2007invention,powell1996interorganizational,makarem2016social,sorenson2001syndication,eagle2010network}.

\begin{figure}[htbp]
   \centering
   \includegraphics[width=.50\textwidth]{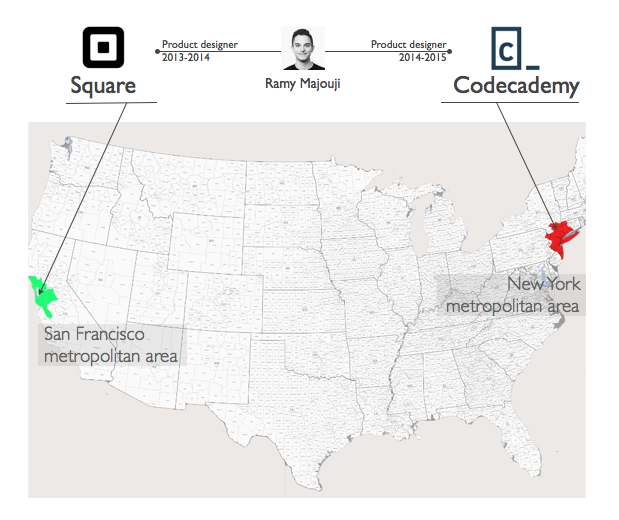} 
   \caption{Example of how the Workforce Mobility Network (WMN) is built. Its 
     nodes represent Metropolitan Statistical Areas (MSA) in
     the United States,
     while each of its links has a weight equal to the number of employees who worked at the companies located in the two corresponding metropolitan areas. For instance,
     Ramy Majouji, who moved from Square Inc. in the San Francisco area (green) to
     Codecademy in the NewYork area (red), acted as a bridge between
     the two companies and contributed to increase the weight of the
     link between the two areas by one.}
   \label{fig:data}
\end{figure}

This work contributes to fill the gap by drawing on a novel dataset from CrunchBase, an online database containing historical records of the evolution of the worldwide startup ecosystem. From this data, to be able to consider network effects, we built and analyzed the first \textit{Workforce Mobility Network} (WMN), which, unlike previous approaches in the literature, can be updated in real-time. The network's nodes  are metropolitan areas, and its links (edges) are workforce flows between area pairs (an edge weight is equal to the number of professionals who have worked at companies located in both of the two metropolitan areas the edge connects).  Figure~\ref{fig:data} provides an
illustration of the procedure adopted to construct WMN: Mr. Ramy Majouji quits his job at Square Inc., a company based in the San Francisco-Oakland-Hayward metropolitan 
area (green), to then join Codecademy, located in the New
York-Newark-Jersey area (red), thus acting as a bridge between the two areas; ultimately, the link between the ``San Francisco-Oakland-Hayward'' node and the ``New
York-Newark-Jersey'' node has a weight equal to the number of unique
workers bridging the companies located in the two areas.

The opportunity to recombine ideas and access relevant knowledge is
crucial for companies that aim at generating innovation~\cite{parise2015twitter,hargadon1998firms,burt1993social}. The likelihood of a company benefiting from new ideas, know-how, and talents is determined not only by the availability of
these resources within the city where the company is located but also
by the opportunity to absorb them from other cities. As such, we
hypothesized that the \textit{most central areas} in WMN,
rather than the most densely populated ones, are the most innovative. 
To test our hypothesis, we  considered two innovation measures
for each metropolitan area $i$: 
\textit{i)}
the number $\mathcal{S}_i$ of successful startups in 
$i$ (a startup is successful if it either was acquired, did an Initial Public Offering
(IPO), or acquired another startup); and 
\textit{ii)} 
the cumulative acquisition price $\mathcal{A}_i$ of all startups in
 $i$. Differently
from commonly used measures of output such as the number
of granted patents, our measures adapt more dynamically to
the rapidly changing market and better reflect a startup's ability to
translate its innovation potential into immediate and tangible
economic value. In a modern innovation landscape characterized more
and more by digital solutions, global outreach, low barrier to
entries, and extremely fast business developments, the number of
patents might not fully reflect actual levels of innovation. Often patents are used as a defensive tool against `patent
trolling'~\cite{cohen2016growing} or are used to discourage the entry of market newcomers  rather than actually being used to produce and commercialize genuinely 
innovative products~\cite{nicholas2014patents}. For completeness, we 
present empirical results considering  patenting rates as a proxy for innovation as well, and do so in the \emph{Supplementary Material}.

In summary, we measured to which extent  WMN
 -- specifically, the centrality of its nodes --  predicts
innovation performance of cities, measured through $\mathcal{S}_i$ and
$\mathcal{A}_i$, and how those predictions compare to previous models' in the literature.

\section*{Results}

All the following models are based on startups that
were active in the United States in 2010, and on all their historical
information up to that year. For each of the metropolitan areas in which these startups were located, we measured the innovation performance indicators $\mathcal{S}_i$ and $\mathcal{A}_i$ in the [2010-2016] period. 
 
\subsection*{Residual variability of population-based models}
\begin{figure}[htbp] 
   \centering
	\includegraphics[width=0.3\textwidth]{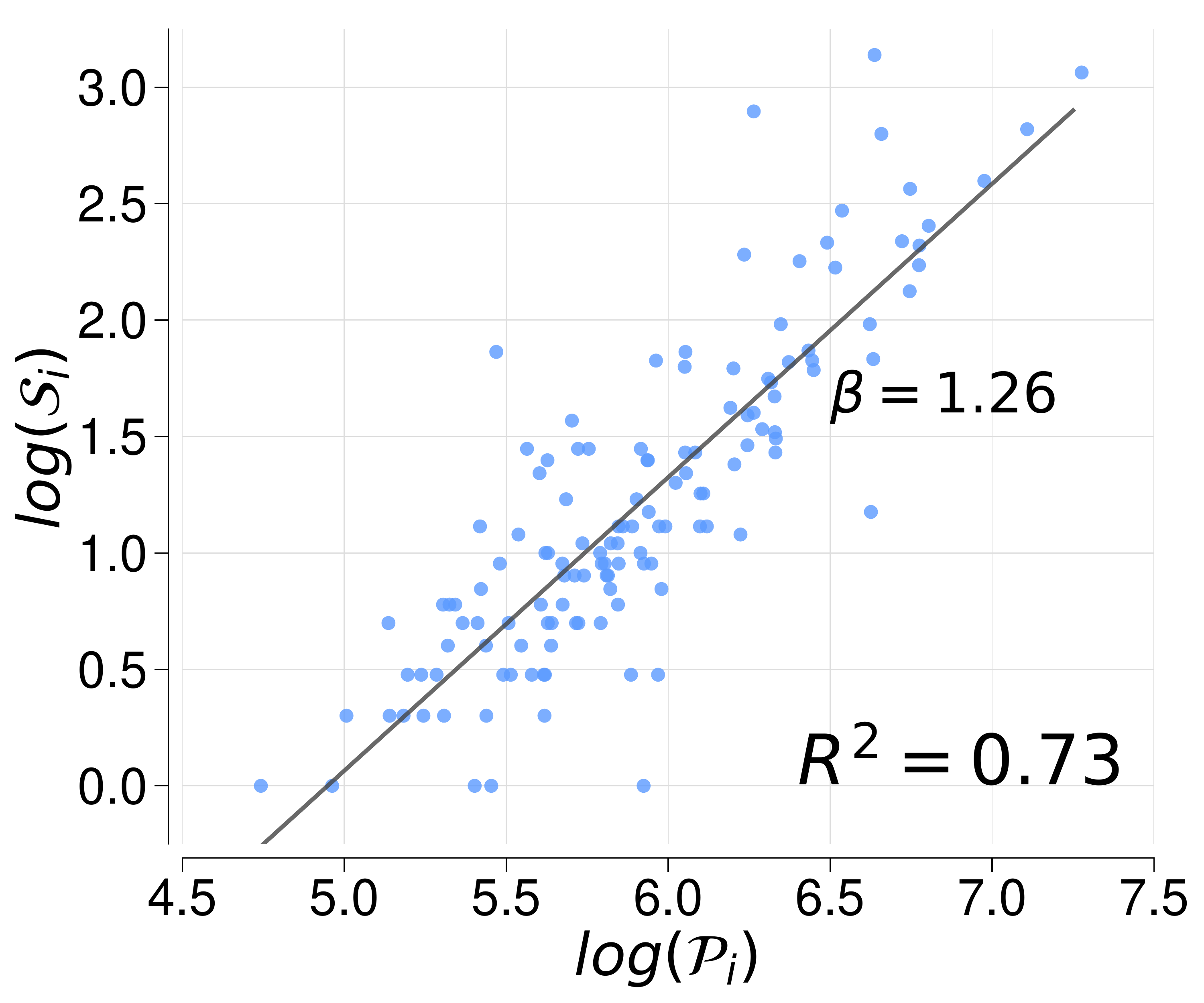} 
	\includegraphics[width=0.3\textwidth]{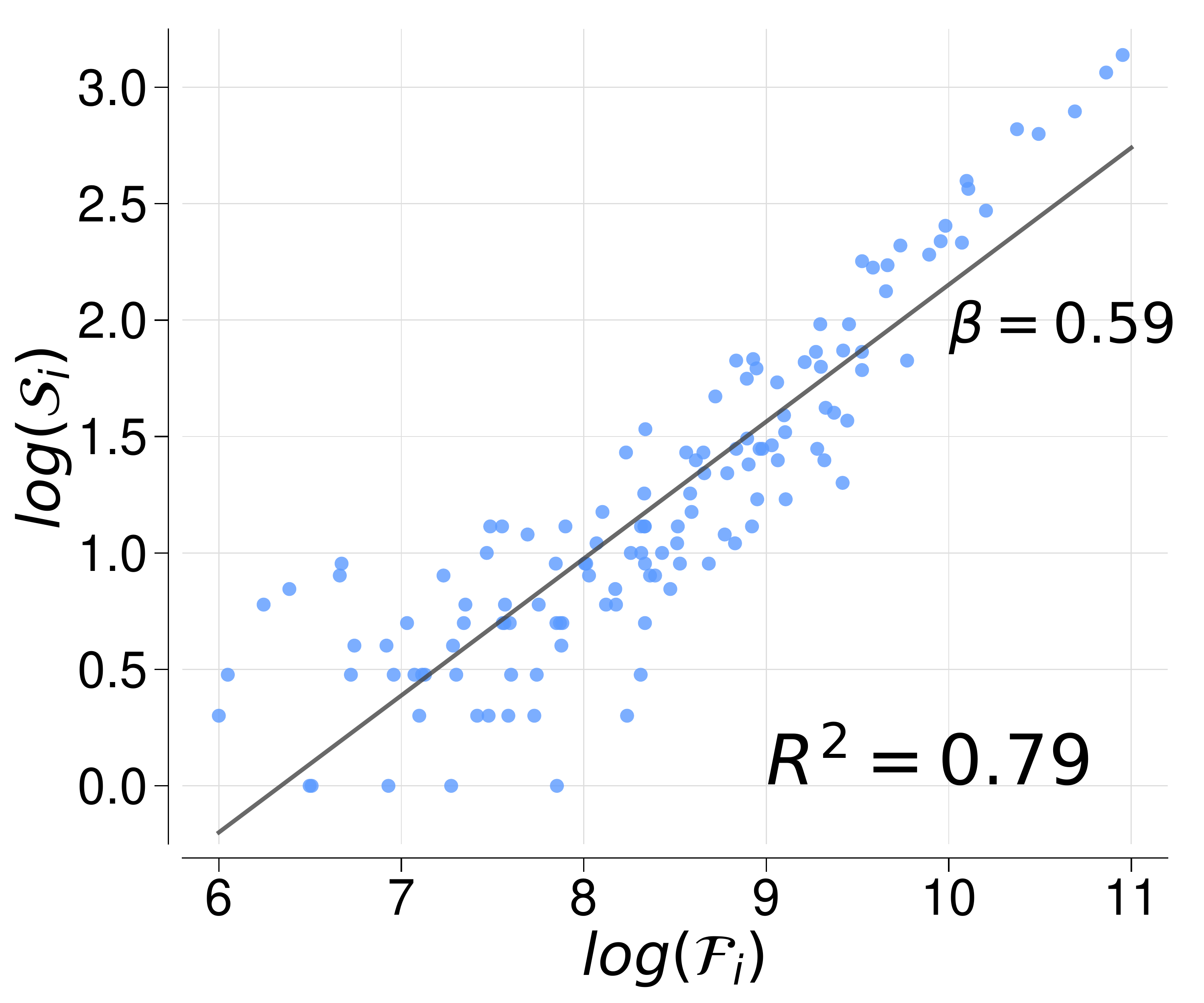}\\
	\includegraphics[width=0.3\textwidth]{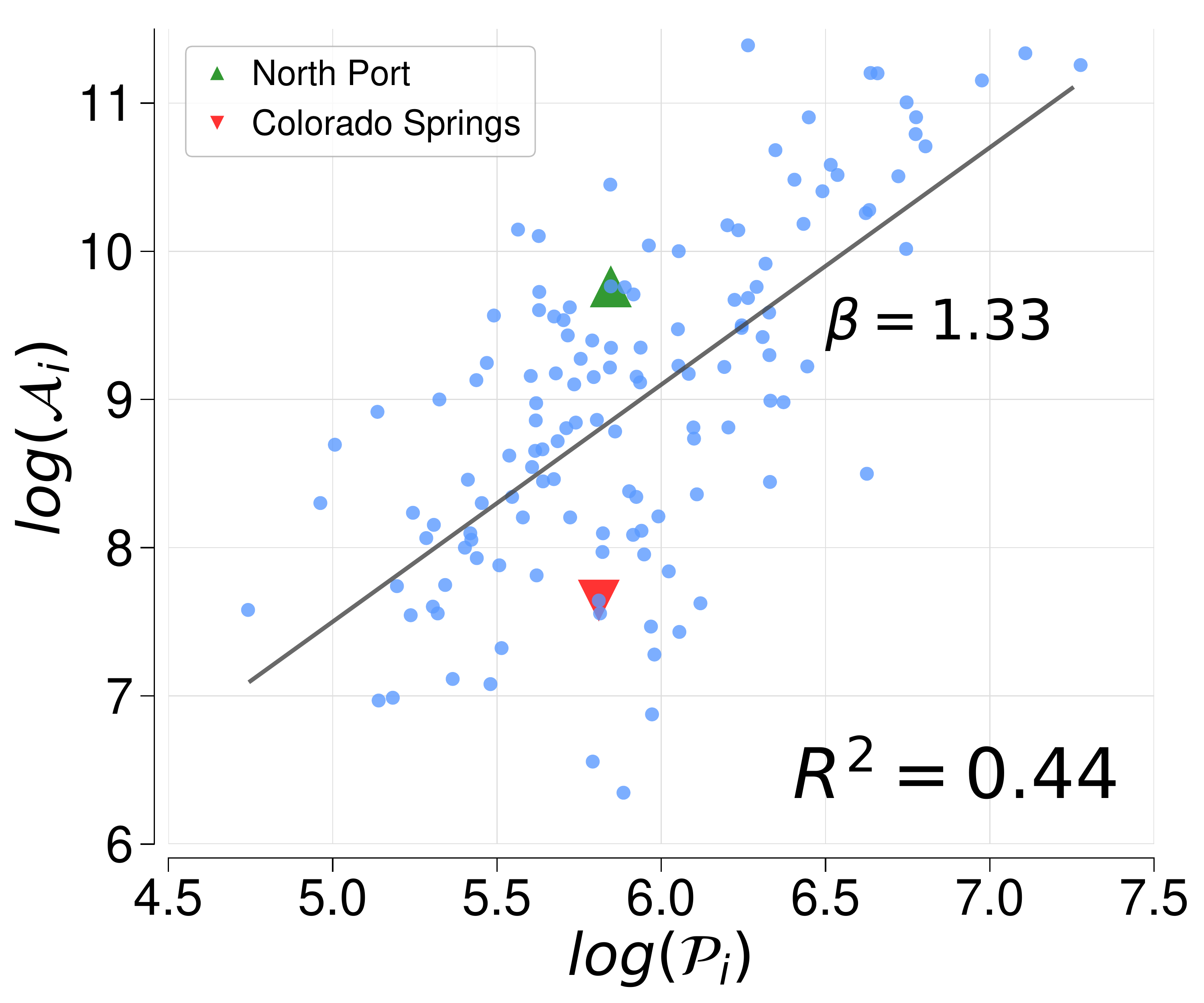}
	\includegraphics[width=0.3\textwidth]{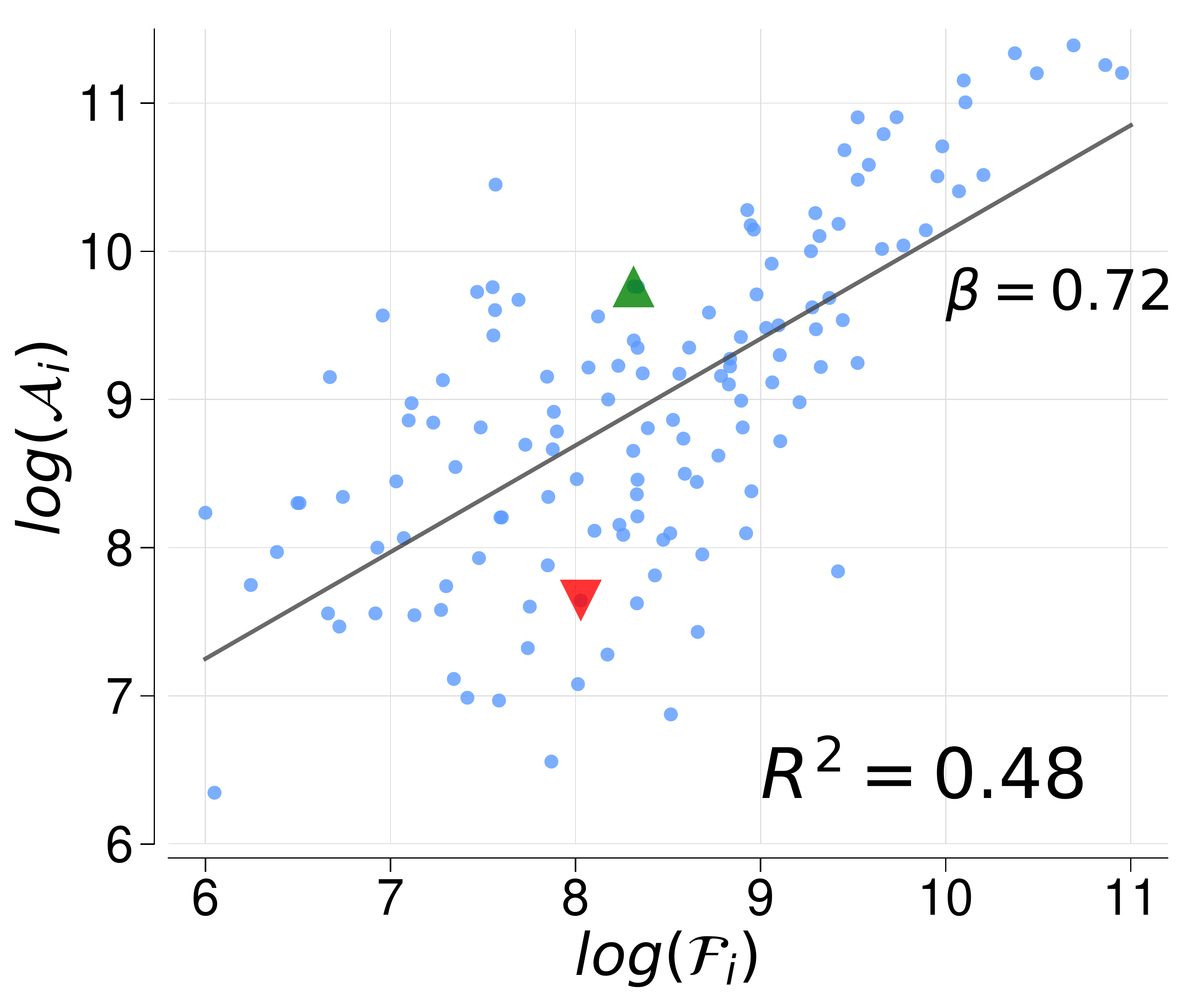} 
        \caption{Scatter plots of our two innovation measures, namely the total number of
          successful startups $\mathcal{S}_i$ and the cumulative
          acquisition price $\mathcal{A}_i$, against population size
          $\mathcal{P}_i$ and total past fundings $\mathcal{F}_i$. Double
          logarithmic plots, coefficients of
          determination $R^2$, and corresponding $\beta$-coefficients for the four least-square linear regressions are shown.}
   \label{fig:population_funding_success_acquisition}
\end{figure}

Consistently with previous work~\cite{bettencourt2007growth}, we found
a non-linear scaling of our two innovation  measures $\mathcal{S}_i$ and $\mathcal{A}_i$ with population size $\mathcal{P}_i$, and with past fundings  $\mathcal{F}_i$ (Figure~\ref{fig:population_funding_success_acquisition}).  

However, despite the correlations being strong, performance variability is still high. Many cities that are similar in size and in past fundings expressed very different performances. For
example, the \textit{North Port-Bradenton-Sarasota} metropolitan area
(Florida) and the \textit{Colorado-Springs} metropolitan area (Colorado) are
very similar with respect to number of startups active in 2010
(respectively $106$ and $99$), population ($\sim 10^6$), and funding
received ($\sim 10^8 \$ $), yet the performances of their companies
are significantly different: companies in ``North
Port-Bradenton-Sarasota'' have been sold for a cumulative value of $5.8
\cdot 10^9 \$ $, while those in ``Colorado-Springs'' reported a cumulative
acquisition price smaller by two orders of magnitude, namely $4.3
\cdot 10^7 \$ $.

Our aim was to investigate to which extent these differences in
performance could be accounted for by other predictors. In particular,
we hypothesized that workforce mobility
explains most of the residual variability.

\subsection*{The Workforce Mobility Network}

We constructed the Workforce Mobility Network (WMN) among metropolitan areas by using CrunchBase job records for all startups active up to the year of 2010. Among the 380
metropolitan areas in the United States, $243$  had at least one active startup in
our data. As a result, the final network had $243$ nodes and $2,169$
edges, and reflected $26,660$ worker flows among metropolitan areas.
The maximum node degree is 165, and the maximum node strength
(the maximum sum of the link weights for a node) is
8,370. The strength distribution follows a power-law function with an
exponent $\sim2$, a value similar to those observed in other
real-world weighted networks~\cite{latora2017}.

\begin{figure}[htbp] 
   \centering
   \includegraphics[width=.49\textwidth]{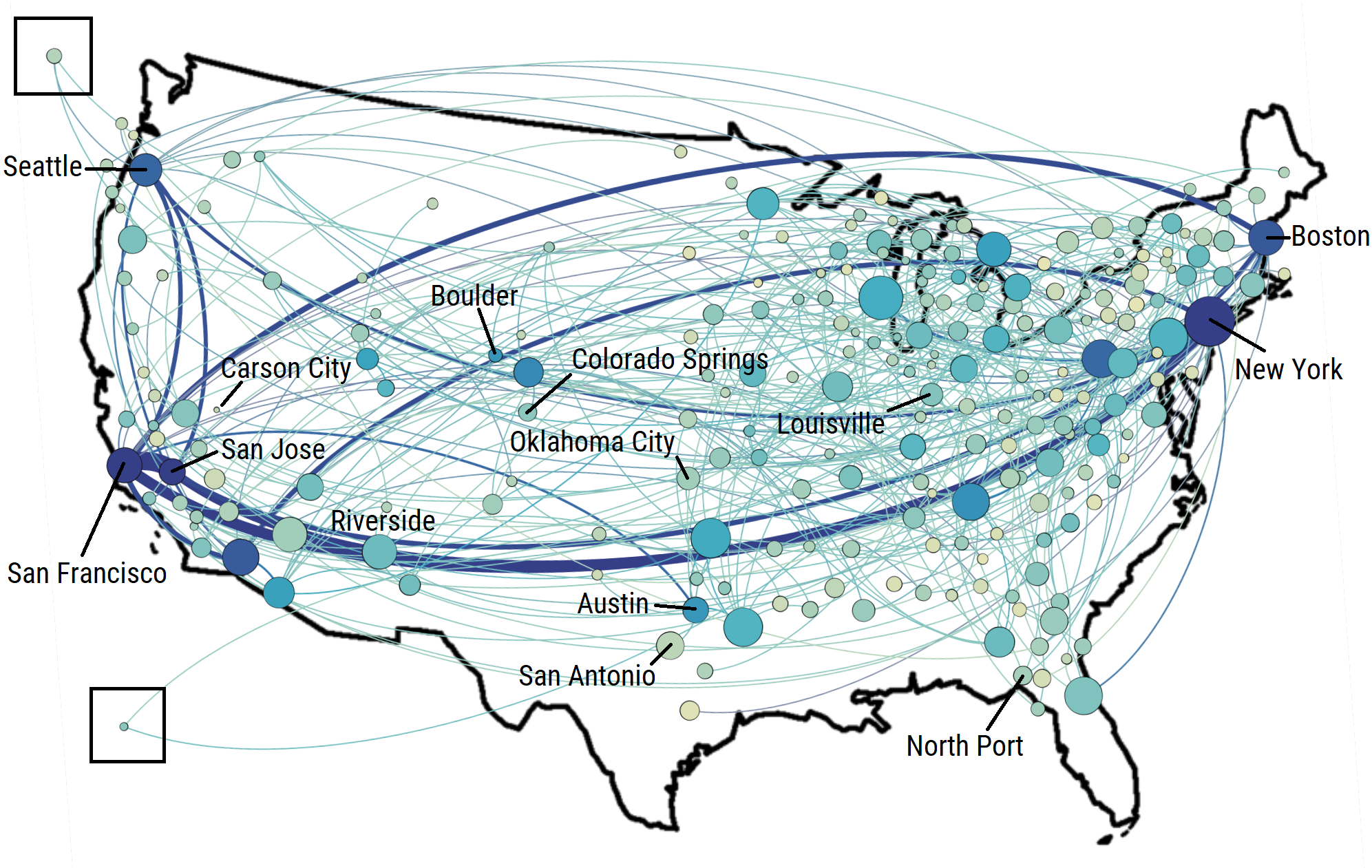} 
	 \includegraphics[width=.49\textwidth]{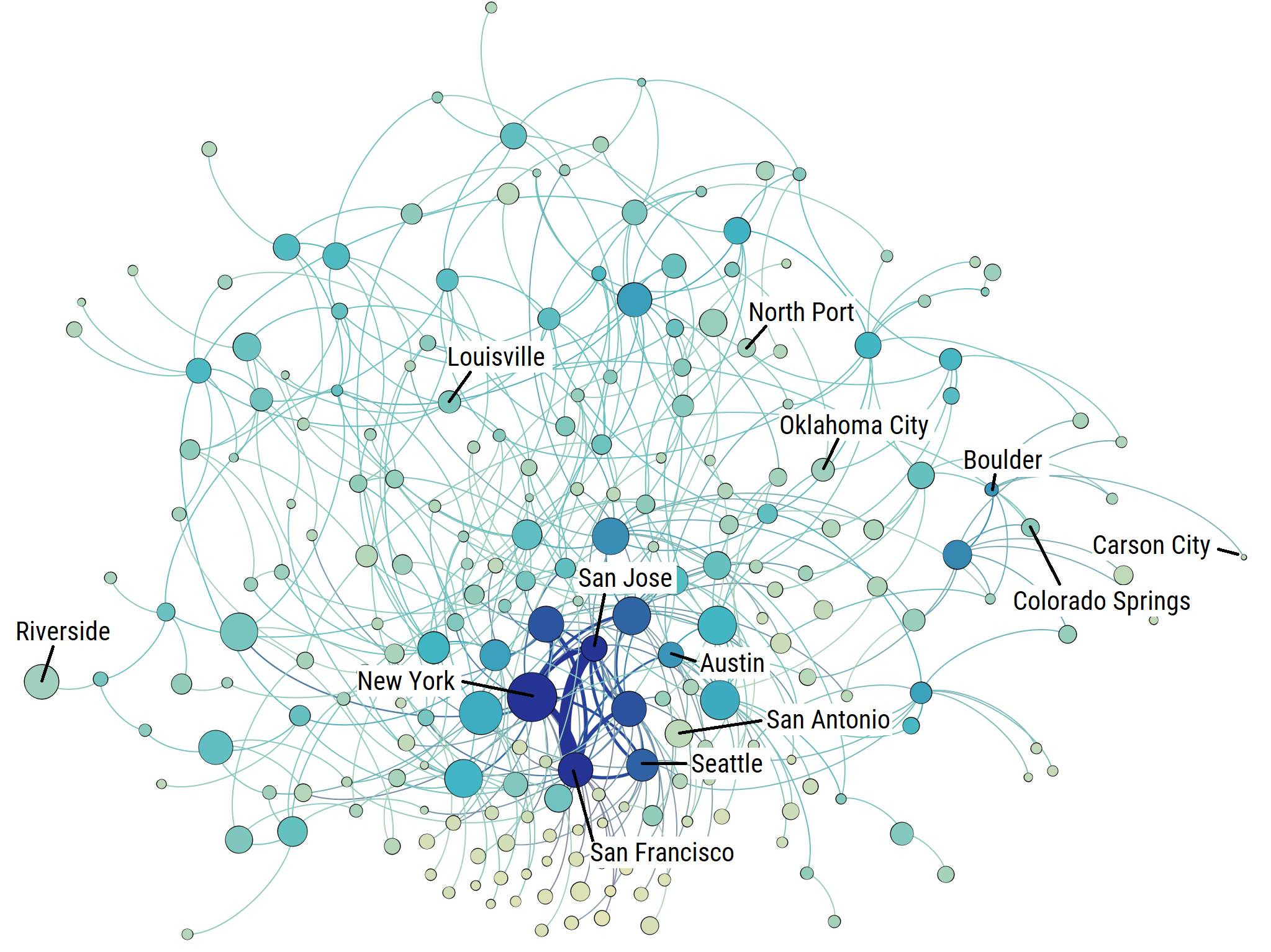} 
   \caption{The backbone of the Workforce Mobility Network (WMN) 
     visualized: (A) on the map of the United States; and (B) on a standard
     force-directed network layout. A node's size is proportional to the area's population size, its color intensity is proportional to its 
     PageRank centrality, and an edge's width is proportional to its weight.}
   \label{fig:GWM}
\end{figure}

To visualize WMN, we projected it onto the map of the
United States, centering its nodes on the metropolitan areas they represent (Figure~\ref{fig:GWM}A). Since the number of edges was high, to improve the visualization, we
reduced the number of displayed edges with a network backbone
extraction algorithm~\cite{coscia2017network}, which identified the most
statistically significant edges for  each node and pruned the rest out. We then  computed each node's centrality according to four measures (see \textit{Methods}), and PageRank yielded the best fit. In Figure~\ref{fig:GWM},  we notice that the most central nodes tend to be US coastal areas, which happen to be linked with each other by the strongest edges. Although population and centrality are in general well
correlated (Spearman rank correlation $\rho=0.70$), large fluctuations
are still observed: indeed, despite being large, several cities do not score high in terms of node centrality  (Figure~\ref{fig:GWM}B).

To identify cities that are small yet central, and viceversa, we ranked 
cities by their ratios $\eta$ between their PageRank centrality
 values $C_i^{PR}$ 
and their population sizes $\mathcal{P}_i$:
\begin{equation}
\eta = \frac{C_i^{PR} / \sum_j{C_j^{PR}} }{\mathcal{P}_i / \sum_j{\mathcal{P}_j}}.
\label{eq:eta}
\end{equation}
Both centrality values and population sizes are normalized by their sums across all areas. Table~\ref{table:topbottom} shows the 10 metropolitan areas with the highest values of
$\eta$, and the 10 with the lowest values.  
Metropolitan areas at the top have higher centrality relative to
their population size. These include large and central
areas such as San Francisco as well as much smaller cities (e.g., Boulder and Ithaca) that are
 remarkably central despite their limited size. On the other hand,
the ten cities at the bottom  are very populous yet not
central in workforce flows, and, with the exception of Virginia Beach, the remaining nine  cities experience relatively limited financial returns from innovation. These findings seem to suggest that
network centrality might predict innovation performance better than what population counts would do. We set out to test that proposition next. 

{\tiny
\begin{table}[!t]
\centering 
\begin{tabular}{lccc|lccc}
 \multicolumn{4}{c|}{Top 10 according to $\eta$} & \multicolumn{4}{c}{Bottom 10 according to $\eta$}\\ 
\hline
 MSA & Population & Price $\mathcal{A}_i$ & $\eta$ & MSA & Population & Price $\mathcal{A}_i$ & $\eta$  \\
\hline
 San Jose CA      & 1.8M & 245B & 8.20 & Riverside CA      & 4.2M & 315M & 0.161 \\
 San Francisco CA & 4.3M & 160B & 6.74 & Oklahoma City OK  & 1.2M & 648M & 0.262 \\
 Boulder CO       & 0.3M & 1.7B & 5.14 & Fresno CA         & 1.0M &  30M & 0.445 \\
 Boston MA        & 4.5M & 159B & 4.82 & Virginia Beach VA & 1.7M & 4.7B & 0.446 \\
 Springfield MO   & 0.4M & 0.3B & 4.73 & Louisville KY     & 1.3M & 229M & 0.464 \\
 Cedar Rapids IA  & 0.3M & 0.3B & 4.42 & Buffalo NY        & 1.1M &  27M & 0.468 \\
 Carson City NV   & 0.1M &  38M & 3.92 & San Antonio TX    & 2.1M & 278M & 0.469 \\
 Seattle WA       & 3.4M &  32B & 3.88 & Columbia SC       & 0.8M &   2M & 0.470 \\
 Austin TX        & 1.7M &  14B & 3.77 & Birmingham AL     & 1.1M & 169M & 0.471 \\
 Ithaca NY        & 0.1M & 0.5B & 3.71 & Sacramento CA     & 2.2M & 981M & 0.489 \\
\end{tabular} 
\caption{Cumulative acquisition prices $\mathcal{A}_i$ for the ten metropolitan areas (MSAs) with the highest values of centrality/population ratio $\eta$, and the ten cities with the lowest values.\label{table:topbottom}}
\end{table} 
}

\subsection*{Predicting innovation performance of cities}

We used linear regression to evaluate the impact of demographic
characteristics and network characteristics on the performance of an area's startups. Linear regression is an approach for modeling a
linear relationship between a dependent variable (our innovation measure $\mathcal{S}_i$
or $\mathcal{A}_i$)
and a set of independent variables, and it does so by associating a so-called
\textit{$\beta$-coefficient} with each independent variable such as the
sum of all independent variables multiplied by their respective $\beta$-coefficients
approximates the value of the dependent variable with minimal
error. Specifically, we used an ordinary least square (OLS) regression
model to estimate the coefficients such that the sum of the squared
residuals between the estimation and the actual value is minimized.
In line with what discussed by Bettencourt \emph{et
al.}~\cite{bettencourt2010urban}, it is more appropriate to express the
dependent variable using absolute values (i.e., number of successful startups, total acquisition
prices) rather than using ratios (e.g., percentage of
successful startups) or per-capita values. That is because these two latter quantities  implicitly assume that the dependent variable (e.g., innovation measure) linearly increases with the independent
variables (e.g., number of existing startups, population size), while we know that it tends to super-linearly increase with them. 

In the regression models, we experimented with two different groups of
predictors: \emph{i)} socio-economic indicators; and \emph{ii)} indicators based on WMN's structure. First, the socio-economic indicators based on the literature are population size~\cite{bettencourt2007growth}, population
density~\cite{jacobs1961death}, and number of patents
granted in each metropolitan area 
up to 2010~\cite{bettencourt2007invention}. To those three indicators, we added two others derived from CrunchBase: the number of active
startups $\mathcal{N}_i$ in 2010, and the total past funding
$\mathcal{F}_i$ raised up to the year of 2010. The number of active startups $\mathcal{N}_i$ is an upper bound for the number of successful ones and, as such, represents an important variable to control for; on the other hand, the independent variable of past funding $\mathcal{F}_i$ is not necessarily correlated with our dependent variable (i.e., with the actual innovation levels of  companies), can be influenced by
factors such as local tax policies, and, as such, can be regarded as a proxy for innovation incentives each area tends to enjoy. 

Second, the indicators based on WMN's structure aim at capturing each area's centrality in the flows of ideas, techniques, knowledge,
creative inputs, and business opportunities~\cite{bon19}. To characterize the 
potential exposure of a metropolitan area to these flows, we computed four centrality measures:
degree centrality, node strength, Google PageRank, and harmonic closeness 
(see \textit{Methods}).  
The node degree and the node strength are local measures that estimate
the potential of an area to exchange resources with its immediate
neighbors, while the latter two are global measures.
The PageRank score of a node is proportional to
the number of ``random walkers'' who happen to traverse that node at stationary state\cite{page1999PageRank}. 
If we imagine knowledge as a collection of discrete units and
assume that these units randomly flow  in WMN, then an area's PageRank score is the fraction of the global knowledge the area has potential access to (e.g., if the score is 0.2, then 20\% of the global knowledge is potentially accessible by the area). In a similar way, area $i$'s
harmonic closeness is  the distance
(measured as the weighted number of hops) that a given unit of
information needs to traverse to reach node $i$ starting from any other
node~\cite{boldi2014axioms,marchiori2000harmony,crucitti2006,pan2011path}. 

Table~\ref{table:regression_success} reports the adjusted coefficients of
determination $R^2$ and
the $\beta$-coefficients for the ten
models. The first 9 models consider  the independent variables separately. We see that predicting acquisition prices $\mathcal{A}_i$
is harder than predicting the number of successful startups
$\mathcal{S}_i$, yet the relative powers of the predictors are mostly
consistent across the two innovation measures.  
All the socio-economic indicators (models
1--5) are good predictors, and, among them, the control variable of the number of active
startups (5) is the most powerful, with $R^2=0.96$ for
$\mathcal{S}_i$, and $R^2=0.76$ for $\mathcal{A}_i$, not least because the
number of active startups is an upper bound for the number of
successful ones. In line with previous empirical
findings~\cite{bettencourt2007growth}, population (1) is positively correlated with both innovation measures. However,  population density (2) is less so. Past fundings (3)
and number of patents (4) are also positively associated, yet have the smallest
$\beta$-coefficients. The last four models (models 6--9)  test our four network centrality measures: PageRank (6) and node strength (7) have higher $\beta$-coefficients
and $R^2$ compared to node degree (8) and
harmonic centrality (9), which do not account for network weights (that is, for the extent to which global knowledge tends to be accessible by each area). Overall, PageRank 
outperforms population by $12\%$ when predicting $\mathcal{S}_i$, and
by $27\%$ when predicting $\mathcal{A}_i$. 

To further disentangle the unique contribution of each predictor, we used 
a stepwise feature selection procedure (see \textit{Methods}
for details) to select the combination of predictors with 
the highest $R^2$. The two models that consist of the selected variables are reported
in the last columns  of the two panels in Table~\ref{table:regression_success} (columns 10). PageRank is the only network metric retained by the
feature selection method because it is the only one that, in combination
with the socio-economic features, improves the overall
prediction. Also, the $\beta$-coefficient of PageRank is the
highest for $\mathcal{A}_i$, and the 
second highest (only after the control variable of the number of active startups)
for $\mathcal{S}_i$. In both cases, the
coefficients of determination are significantly larger than those
obtained for the other variables, especially than those obtained for 
population size and density.

\begin{figure}[htbp]
   \centering
   \includegraphics[width=0.35\textwidth]{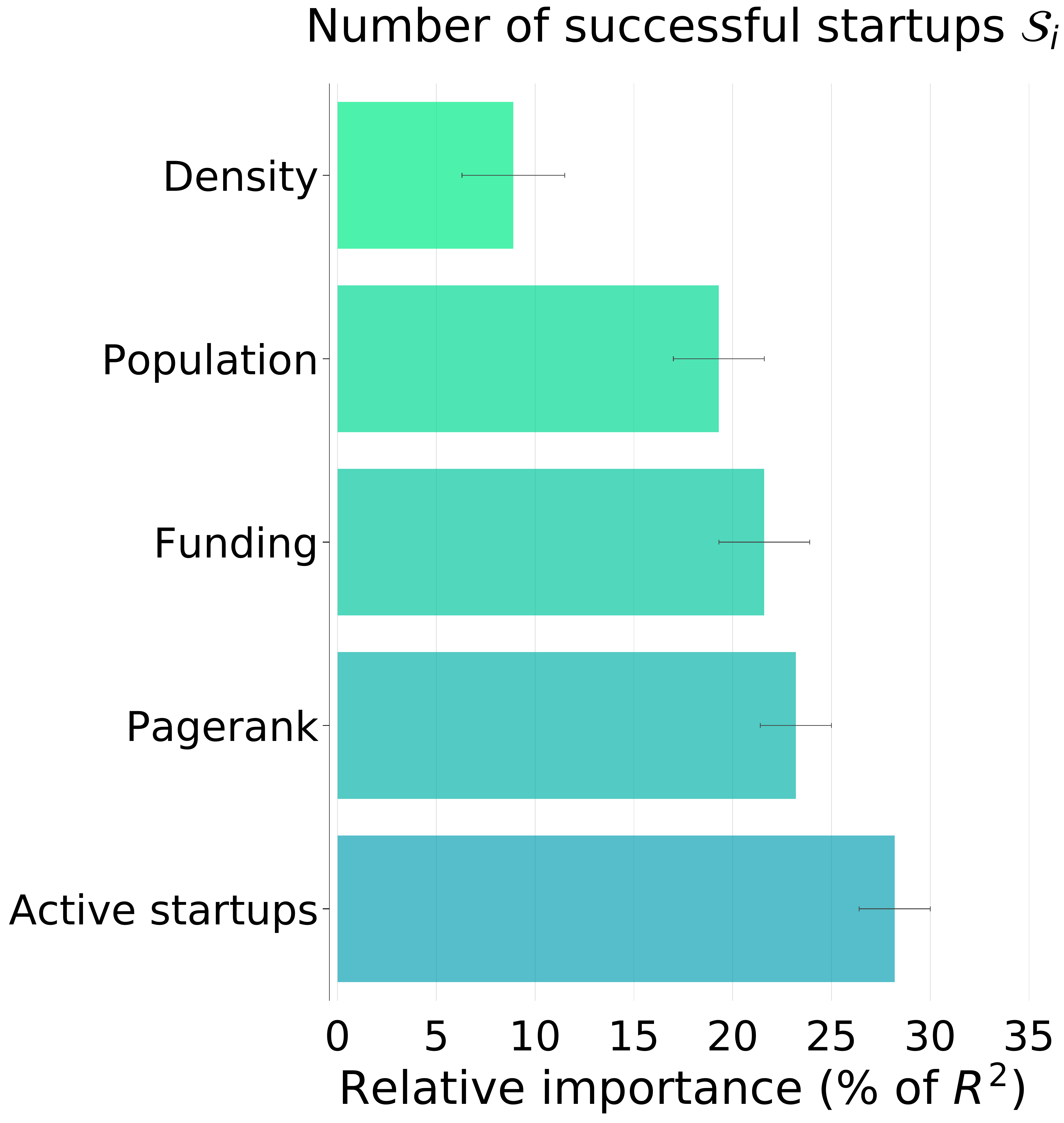}
	\includegraphics[width=0.35\textwidth]{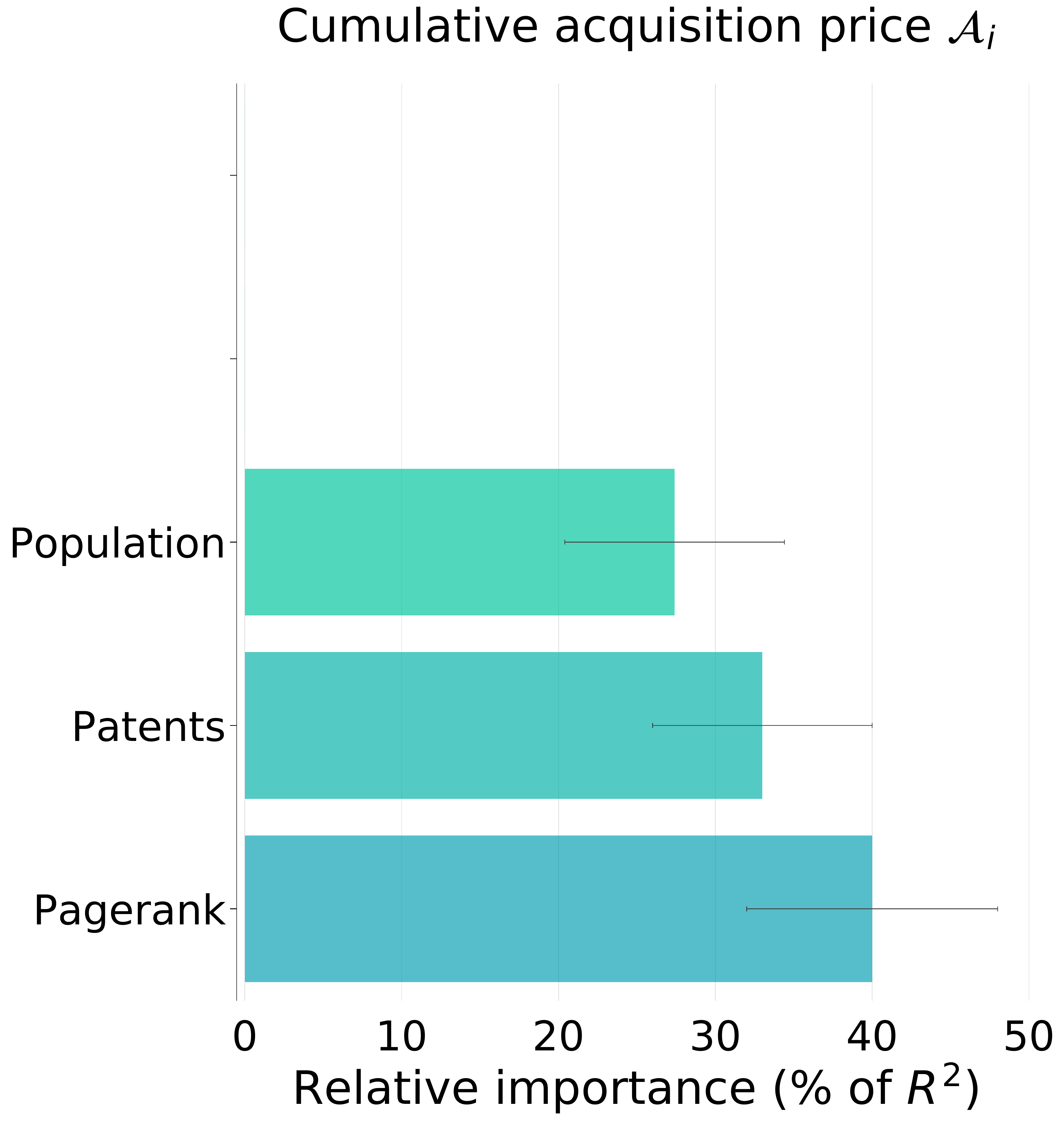}
   \caption{Relative contribution of each variable to the total
     variance explained by the regression models in
     Table~\ref{table:regression_success}. The relative contribution is expressed
     as percentage of the $R^2$ determined by each variable.}
   \label{fig:relative_importance}
\end{figure}

In multivariate regressions, if
the independent variables are perfectly independent, then 
the coefficient of determination $R^2$
decomposes itself into the sum of the squares of the Pearson's
correlation coefficients computed for each variable separately.
However, in our case, as in the majority or real-world scenarios, most of the
variables are correlated with each other, and the sum of each
independent $R^2$ exceeds the one obtained for the multivariate
regression (model 10). To properly decompose the relative contribution
of the correlated independent variables, we used
the LGM method~\cite{lindeman1980introduction} and computed the relative importance of each predictor (Figure~\ref{fig:relative_importance}). Interestingly, after controlling for the number of active startups, PageRank is confirmed to be the predictor that explains most of the variability in
the data.

{\tiny
\begin{table}[!htbp] \centering 
\setlength{\tabcolsep}{1pt}
\begin{tabular}{@{\extracolsep{5pt}}lcccccccccc} 

\\[-1.8ex]\hline 
\hline \\[-1.8ex] 
 & \multicolumn{10}{c}{\textit{Dependent variable:} number of successful startups $\mathcal{S}_i$} \\ 
 \\[-1.8ex] & (1) & (2) & (3) & (4) & (5) & (6) & (7) & (8) & (9) & (10)\\ 
\hline \\[-1.8ex] 
 Population          & 0.853$^{***}$ &  &  &  &  &  &  &  &  & 0.087$^{*}$ \\ 
                     & (0.046)       &  &  &  &  &  &  &  &  & (0.050) \\ 
 Pop. density        &  & 0.627$^{***}$ &  &  &  &  &  &  &  & 0.050$^{*}$ \\ 
                     &  & (0.068) &  &  &  &  &  &  &  & (0.028) \\ 
 Funding raised      &  &  & 0.890$^{***}$ &  &  &  &  &  &  & 0.164$^{***}$ \\ 
                     &  &  & (0.040) &  &  &  &  &  &  & (0.048) \\ 
 Patents             &  &  &  & 0.862$^{***}$ &  &  &  &  &  &  \\ 
                     &  &  &  & (0.044) &  &  &  &  &  &  \\ 
 Active startups     &  &  &  &  & 0.961$^{***}$ &  &  &  &  & 0.511$^{***}$ \\ 
                     &  &  &  &  & (0.024) &  &  &  &  &  (0.084) \\ 
 Network PageRank    &  &  &  &  &  & 0.907$^{***}$ &  &  &  & 0.222$^{***}$ \\ 
                     &  &  &  &  &  & (0.037) &  &  &  & (0.049) \\ 
 Network strength    &  &  &  &  &  &  &  0.911$^{***}$ &  &  &  \\ 
                     &  &  &  &  &  &  &  (0.036) &  &  &  \\ 
 Network degree      &  &  &  &  &  &  &  & 0.871$^{***}$ &  &  \\ 
                     &  &  &  &  &  &  &  & (0.043) &  &  \\ 
 Harmonic centrality &  &  &  &  &  &  &  &  & 0.751$^{***}$ &  \\ 
                     &  &  &  &  &  &  &  &  & (0.058) &  \\ 
 Constant            & $-$0.000 & 0.000 & 0.000 & $-$0.000 & $-$0.000 & $-$0.000 & $-$0.000 & $-$0.000 & $-$0.000 \\ 
                     & (0.045) & (0.068) & (0.040) & (0.044) & (0.024) & (0.037) & (0.036) & (0.043) & (0.057) & (0.021) \\ 
\hline \\[-1.8ex] 
Adjusted R$^{2}$     & 0.73 & 0.39 & 0.79 & 0.74 & 0.92 & 0.82 & 0.83 & 0.76 & 0.56 & 0.94  \\
\end{tabular} 

\begin{tabular}{@{\extracolsep{5pt}}lcccccccccc} 
\\[-1.8ex]\hline 
\hline \\[-1.8ex] 
\\[-1.8ex]  & \multicolumn{10}{c}{ \textit{Dependent variable:} cumulative acquisitions prices $\mathcal{A}_i$} \\ 
\\[-1.8ex] & (1) & (2) & (3) & (4) & (5) & (6) & (7) & (8) & (9) & (10)\\ 
\hline \\[-1.8ex]
 Population          & 0.668$^{***}$ &  &  &  &  &  &  &  &  & 0.153$^{*}$ \\ 
                     & (0.065) &  &  &  &  &  &  &  &  & (0.091) \\ 
 Pop. density        &  & 0.471$^{***}$ &  &  &  &  &  &  &  &  \\ 
                     &  & (0.077) &  &  &  &  &  &  &  &  \\ 
 Funding raised      &  &  & 0.698$^{***}$ &  &  &  &  &  &  &  \\ 
                     &  &  & (0.063) &  &  &  &  &  &  &  \\ 
 Patents             &  &  &  & 0.715$^{***}$ &  &  &  &  &  & 0.254$^{**}$ \\ 
                     &  &  &  & (0.061) &  &  &  &  &  & (0.101) \\ 
 Active startups     &  &  &  &  & 0.758$^{***}$ &  &  &  &  &  \\ 
                     &  &  &  &  & (0.057) &  &  &  &  &  \\ 
 Network PageRank    &  &  &  &  &  & 0.748$^{***}$ &  &  &  & 0.431$^{***}$ \\ 
                     &  &  &  &  &  & (0.058) &  &  &  & (0.098) \\ 
 Network strength    &  &  &  &  &  &  & 0.719$^{***}$ &  &  &  \\ 
                     &  &  &  &  &  &  & (0.061) &  &  &  \\ 
 Network degree      &  &  &  &  &  &  &  & 0.673$^{***}$ &  &  \\ 
                     &  &  &  &  &  &  &  & (0.065) &  &  \\ 
 Harmonic centrality &  &  &  &  &  &  &  &  & 0.587$^{***}$ &  \\ 
                     &  &  &  &  &  & &  &  & (0.071) &  \\ 
 Constant            & 0.000 & 0.000 & 0.000 & 0.000 & 0.000 & 0.000 & 0.000 & 0.000 & 0.000 & 0.000 \\ 
                     & (0.065) & (0.077) & (0.062) & (0.061) & (0.057) & (0.058) & (0.060) & (0.064) & (0.070) & (0.055) \\ 
\hline \\[-1.8ex] 
Adjusted R$^{2}$     & 0.44 & 0.22 & 0.48 & 0.51 & 0.57 & 0.56 & 0.51 & 0.45 & 0.34 & 0.6  \\
\hline 
\hline \\[-1.8ex] 
\textit{}  & \multicolumn{10}{r}{$^{*}$p$<$0.1; $^{**}$p$<$0.05; $^{***}$p$<$0.01} \\ 
\end{tabular}
\caption{Standardized $\beta$-coefficients of the regression models to
  predict the two dependent variables of innovation performance, namely 
  the number of successful startups
  $\mathcal{S}_i$ and the cumulative acquisition price $\mathcal{A}_i$. Standard errors for the coefficients are reported in
  parenthesis.}
\label{table:regression_success}
\end{table} 
}

\section*{Discussion} 

This is the first study that has built a Workforce Mobility  Network  at the scale
of an entire country from open data, and that has  shown that this network's structural characteristics  are  powerful  predictors of urban innovation. It turns out that startups benefit from being located in areas that are central in workforce flows: indeed, global network measures tend to predict long-term innovation better than even what cumulative investments do.

Our study comes with limitations that are mostly determined by our  data. No sufficient longitudinal data was available for testing causal relationships. Furthermore, startups do not have to publicly disclose their funding rounds or acquisition prices:  83\% of the funding rounds in our dataset, for example, have been fully disclosed on CrunchBase. Yet, being of random nature, such missing data has little impact on our two innovation measures, and no impact on a comparative evaluation of areas.

\section*{Methods}

\subsection*{Datasets}

We combined data from three sources. First, from US census data, we extracted information about population size, land area, and population density at the level of \textit{Metropolitan Statistical Area} (MSA). Second, from the United States Patent and Trademark Office (USPTO), we associated  the numbers of patents granted in the year of 2010 with the inventors'  metropolitan areas. Third, from the CrunchBase web APIs, we collected all information regarding \textit{organizations} and \textit{people} (workers). For each organization we extracted data on: address of the headquarter, foundation date, funding rounds, acquisitions (also referred to as \textit{exits}), initial public offers (IPOs), status (active, closed), and team members. The address, in turn, consists of street name, zip-code, city name, and state. Funding rounds record the financial investment of individuals or venture capital firms into a company (organization), i.e., the purchase of a certain percentage of ownership of the company, while acquisitions indicate the transfer of the company's total ownership to another company. The data on funding rounds and acquisitions include the parties involved, the date, and the monetary value of the transaction in US dollars. We were able to associate the companies in our data with 369 (out of the 374) metropolitan areas. Workers are linked to organizations through the professional roles they hold. Examples of role titles are \textit{CEO}, \textit{founder}, \textit{board member}, and \textit{employee}.
Workers can have multiple jobs/roles within the same organization or across different organizations. About 42\% of all the job records include a starting date allowing for a longitudinal analysis of the flow of workers between various firms. 

\subsection*{Centrality measures}

Different measures of centrality have been proposed over the years to
quantify the importance of a node in a complex network~\cite{latora2017}.
In this work, we computed four centrality measures for each WMN
node: degree centrality, node strength, harmonic closeness, and Google PageRank.

Let $G$ be a weighted graph with $N$ nodes described by the $N \times N$ weighted
adjacency matrix $W=\{ w_{ij} \}$ whose entry $w_{ij}$ is equal to the
weight of the link connecting node $i$ to node $j$, or is equal to $0$ if the
nodes $i$ and $j$ are not connected. 
As for the case of $G$ being an unweighted graph, we  define the
adjacency matrix $A=\{ a_{ij} \}$ of  $G$, which simply indicates which pairs of nodes are connected with  a $N \times N$ matrix such that $a_{ij}=1$ if $w_{ij} \neq 0$,
and $a_{ij}=0$ if $w_{ij} = 0$. 
\\
Our first centrality measure out of the four is {\em degree centrality}, which is based on 
the idea that important nodes are those with the largest number of ties 
to other nodes in the graph. The degree centrality of node $i$ is
defined as: 
\begin{equation}
\label{DC}
C^D_i = \frac{k_i} {N-1} 
       = \frac{  \sum_{ j= {1}}^N   a_{ij}   } {N-1}
\end{equation}
where $k_i$ is the degree of node $i$. 
\\
Our second centrality measure is  {\em strength centrality}. For each node $i$, this is defined as: 
\begin{equation}
\label{SC}
C^S_i = \frac{s_i} { \sum_j s_j }   
       = \frac{  \sum_{j= {1}}^N   w_{ij}   } { \sum_{i,j}   w_{ij}   }
\end{equation}
where strength $s_i$ of node $i$ is the sum of the weights of the edges
incident in $i$.
\\
Our third centrality measure is the {\em harmonic closeness centrality}~\cite{marchiori2000harmony}. For each node $i$, it measures its minimum distance or  
geodesic $d_{ij}$ to another node $j$, i.e., the minimum  number of edges  
traversed to get from $i$ to $j$: 
\begin{equation} 
\label{CC}
C^C_i = \sum_{j}\frac{1}{d_{ij}}.
\end{equation}

Our fourth and final centrality measure is the {\em PageRank centrality}. For each  node $i$, this is  the stationary probability that a  ``surfer'' that randomly travels on the network's links arrives at node $i$. It is recursively defined as:
\begin{equation} 
\label{PRC}
C^{PR}_i = \frac{1-\alpha}{N} + \alpha \cdot \sum_{\{j | (i,j) \in A\}} \frac{C^{PR}_j}{k_j}
\end{equation}
where $k_j$ is the degree of node $j$, and $\alpha$ is a \emph{damping factor} (traditionally set to 0.85) that models the probability of the surfer following an existing link instead of jumping to any other node picked at random with uniform probability. In this work, we considered a weighted version~\cite{xing2004weighted} of the PageRank centrality that sets the probability of following a link proportional to the weight of that link. Formally, this is expressed as:
\begin{equation} 
\label{PRC}
C^{PR}_i = \frac{1-\alpha}{N} + \alpha \cdot \sum_{\{j | (i,j) \in A\}} C^{PR}_j \cdot \frac{w_{ji}}{s_j}
\end{equation}
where the factor $\frac{w_{ji}}{s_j}$ expresses the probability of transitioning from node $j$ to node $i$ being equal to the weight of the link between $j$ and $i$ ($w_{ji}$) divided by the total strength of node $j$ ($s_j$). The PageRank values are computed with an iterative procedure (implemented efficiently through the so-called power method~\cite{arasu2002pagerank}) that starts by assigning a uniform PageRank value to all nodes $C^{PR}_i = 1/N$, and runs until convergence.

For all the four centrality measures, we considered their normalized versions $\hat{C_i} = \frac{C_i}{\sum_{j=1}^{N} C_j}$ such that the sum of centrality scores over all the nodes in the network is equal to 1.

\subsection*{Regression}

Since all regression variables had skewed distributions, we log-transformed them using base-10 logarithm, with the only exception of the harmonic closeness centrality. To appropriately compare the $\beta$-coefficients, we standardized all regression variables. Given a regressor $x$, its standardised version $\tilde x$ is given by: $\tilde x = (x-\mu) / \sigma $, where $\mu$ is the average value of the variable $x$ in the sample, and $\sigma$ its standard deviation. Feature selection in the last model (model 10) was performed with the feature selection procedure \textit{stepAIC} implemented in the R standard packages. To estimate the relative feature importance, we used the implementation of the LGM method provided in R in the package \textit{relaimpo}~\cite{gromping2006relative}.

\section*{Data Availability}

All the datasets used in this work can be fully and freely downloaded from the Web. The CrunchBase data is available through its public API at \url{https://data.CrunchBase.com}, patent data can be downloaded from \url{http://www.patentsview.org/download}, and US census data from \url{https://www.census.gov}. To map CrunchBase firms to metropolitan areas, we used the census data available here: \url{https://www.census.gov/geo/maps-data/data/relationship.html}.

\section*{Supplementary information}

\subsection*{Patents granted as outcome variable in the regressions}

We constructed several linear regression models to assess the relative contribution of various predictors (population, number of patents, population density, network metrics) in explaining the variability in city's innovation performance. As a measure of city performance (dependent variable) we have proposed two indicators that directly account for the economic value produced by the start-up firms based in a given city. To compare our results more directly with previous work~\cite{bettencourt2007invention}, we also considered patents granted in year 2010 as alternative performance measure (Tab.~\ref{table:regression_success_patents}). Also in this case, network features are better predictors than population or population density. In this case, the feature selection indicates network strength as the strongest network predictor which, jointly with funding raised and number of active startups, reaches an $R^2$ of $0.79$.

{\tiny
\begin{table}[!htbp] \centering 
\setlength{\tabcolsep}{1pt}
\begin{tabular}{@{\extracolsep{5pt}}lcccccccc} 
\\[-1.8ex]\hline 
\hline \\[-1.8ex] 
% & \multicolumn{10}{c}{\textit{Dependent variable:}} \\ 
%\cline{2-11} 
\\[-1.8ex]  & \multicolumn{8}{c}{ \textit{Dependent variable:} patents granted} \\ 
\\[-1.8ex] & (1) & (2) & (3) & (4) & (5) & (6) & (7) & (8) \\ 
\hline \\[-1.8ex]
 Population          & 1.129$^{***}$ &  &  &  &  &  &  &     \\ 
                     & (0.084) &  &  &  &  &  &  &     \\ 
 Pop. density        &  & 1.117$^{***}$ &  &  &  &  &  &      \\ 
                     &  & (0.140) &  &  &  &  &  &      \\ 
 Funding raised      &  &  & 0.551$^{***}$ &  &  &  &    & 0.157$^{***}$    \\ 
                     &  &  & (0.032) &  &  &  &  &    (0.058) \\ 
 Active start-ups    &  &  &  & 0.941$^{***}$ &  &  &  &   0.458$^{***}$    \\ 
                     &  &  &  & (0.046) &  &  &  &    (0.125) \\ 
 Network PageRank    &  &  &  &  & 1.062$^{***}$ &  &  &     \\ 
                     &  &  &  &  & (0.070) &  &  &    \\ 
 Network strength    &  &  &  &  &  & 0.693$^{***}$ &  &   0.211$^{***}$     \\ 
                     &  &  &  &  &  & (0.037) &  &    (0.086) \\ 
 Harmonic centrality &  &  &  &  &  &  & 0.984$^{***}$       \\ 
                     &  &  &  &  &  &  & (0.080) &      \\ 
 Constant            & -3.908$^{***}$ & 7.108$^{***}$ & -1.863$^{***}$ & 0.848$^{***}$ & 5.615$^{***}$ & 1.632$^{***}$ & 0.004 & 0.167$^{***}$ \\ 
                     & (0.065) & (0.077) & (0.062) & (0.061) & (0.057) & (0.058) & (0.060) & (0.064)   \\ 
\hline \\[-1.8ex] 
Adjusted R$^{2}$     & 0.58 & 0.32 & 0.69 & 0.76 & 0.63 & 0.73 & 0.53 & 0.79  \\
\hline 
\hline \\[-1.8ex] 
& \multicolumn{8}{r}{$^{*}$p$<$0.1; $^{**}$p$<$0.05; $^{***}$p$<$0.01} \\ 
\end{tabular}
\caption{Standardised beta coefficients of the regression models to predict number of patents granted in 2010. Standard errors of the coefficients are reported in parenthesis.}
\label{table:regression_success_patents}
\end{table}
}

\begin{figure}[htbp]
   \centering
   \includegraphics[width=.5\textwidth]{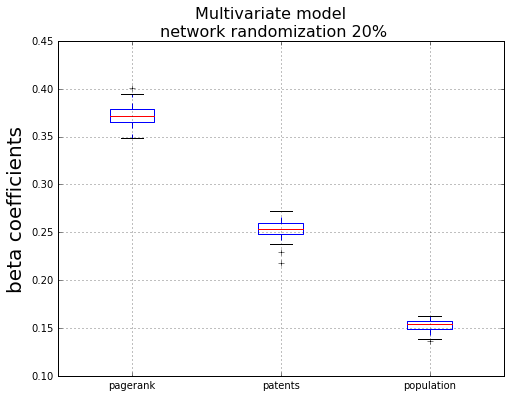} 
   \caption{Regression model to predict the cumulative acquisition price $\mathcal{A}_i$: distributions of beta coefficients over 100 simulations where $20\%$ of the overall link weight is reduced at random.}
   \label{fig:randomisation}
\end{figure}

\section*{Robustness of results to data sparsity}

Even though the CrunchBase dataset is, to date, the largest open dataset that tracks job mobility across start-ups, it does not include the \textit{full} information on mobility. To investigate how much sparsity of the data entries might affect our results and their interpretation, we simulated a scenario where random job entries are dropped and checked whether the relative importance of the predictive variables remain stable. We decreased the overall weight of links in the network by $20\%$, selecting affected links at random and removing them when their strength was reduced to 0. We then run a regression model to predict the cumulative acquisition price $\mathcal{A_i}$ using PageRank, number of patents, and population (the three varibles picked by stepwise feature selection). The distributions of the beta coefficients over 100 random simulations (Fig.~\ref{fig:randomisation}) reveals that the relative weight the coefficient does not change compared to the model that uses all the available data: PageRank has a much higher predictive power than population.

%\bibliography{sample}

\section*{Acknowledgements}
We thank Valerio Ciotti for his help in collecting the data. VL work was funded by the Leverhulme Trust Research Fellowship ``CREATE: the network components of creativity and success''.  

\section*{Author contributions statement}

MB conducted the experiments and analysed the results. All authors conceived the experiments and contributed to write the manuscript.

\end{document}